# A Resolution for Shared Memory Conflict in Multiprocessor System-on-a-Chip

Shaily Mittal[1] and Nitin[2]

[1]Department of Computer Science & Engineering, ITM University, Gurgaon-122017, India

[2]Department of Computer Science & Engineering and Information Technology, Jaypee University of Information Technology, P.O. Waknaghat, Solan-173234, Himachal Pradesh, India

**Abstract.** Now days, manufacturers are focusing on increasing the concurrency in multiprocessor system-on-a-chip (MPSoC) architecture instead of increasing clock speed, for embedded systems. Traditionally lock-based synchronization is provided to support concurrency; as managing locks can be very difficult and error prone. Transactional memories and lock based systems have been extensively used to provide synchronization between multiple processors [1] in general-purpose systems. It has been shown that locks have numerous shortcomings over transactional memory in terms of power consumption, ease of programming and performance. In this paper, we propose a new semaphore scheme for synchronization in shared cache memory in an MPSoC. Moreover, we have evaluated and compared our scheme with locks and transactions in terms of energy consumption and cache miss rate using SimpleScalar functional simulator.

## 1. Introduction

A System-on-a-chip is defined as integration of all components of a computer or some other electronic system into a single integrated circuit (chip). It may contain digital, analog, mixed-signal, and often radio-frequency functions–all on a single chip substrate. Embedded system is a typical area of application of systems on chip. With the expansion in the technologies and requirement of fast and compact systems, need to have more and more electronic circuits on a single chip has increased drastically. A key to this problem leads to evolution of multiprocessor system on a single chip. The multiprocessor System-on-a-chip (MPSoC) [2] is a system-on-a-chip (SoC) which has multiple processors on a single chip, generally used for embedded applications. It is used by platforms that include multiple, generally heterogeneous, processing elements with particular functionalities which contain a memory hierarchy and I/O components. An on-chip interconnect is used to link all these components to each other on the chip such as AMBA interconnect.

Multiple processors on a chip communicate through shared caches [4]. Integrated platforms for embedded applications [3] are even more aggressively pushing core-level parallelism. SoCs with tens of cores are ordinary [5], [6], [7], [8] and platforms with hundreds of cores have been already announced [9]. Multi-core architectures have the advantages of increased power-performance scalability and faster design cycle time by utilizing replication of pre-designed components. If applications make use of a high level of concurrency, then only we can attain power and performance benefits.

One of the toughest challenges to be addressed by multi-core architects is how to expose application parallelism in systems Thread level parallelism brings revolution in MPSoC [10]. As multiple threads can be executed simultaneously, it increases the real advantage of multiple processors on a single chip [11]. However, this also leads to a problem of concurrent access to cache by multiple processors. When more than one processor simultaneously wants to access same shared cache then there is a need of synchronization mechanism.

In this paper, we are presenting the use of semaphores to solve the problem of synchronization among processors. A number of solutions for the problem of cache coherency are already given by many researchers and that we are going to discuss in Section 2. Section 3 describes our proposed solution to the problem with algorithm in detail. Simulation environment is discussed in Section 4 with Section 5 explaining our experimental results of performance evaluation and comparison. Lastly, the work is concluded in Section 6 with tentative future work.

## 2. Background and Motivation

Locks are the most common approach to synchronization [12]. Locked data item in memory by a processor cannot be locked and used by any other processor at the same time. In spite of their wide use, Locks have many disadvantages. Locks also cause vulnerability to thread failures and delays: if a thread holding a lock is delayed by a context switching or page fault, other running threads may be blocked. Locks also hamper concurrency because they must be used conventionally that is a thread must acquire a lock even in rare case of conflict possibility.

Authors in [13] evaluated the energy costs of two approaches to multiprocessor memory synchronization: transactional memory versus locking. The behavior of these synchronization approaches is highly dependent on the system contention level. Hence, their results shows that both transactional memory and standard locking code were not designed with energy consumption in mind, and hence the design of energy-aware synchronization mechanisms remains a largely unexplored area. Moreover, authors suggest a promising energy-aware approach of speculative synchronization for low contention programs and serialization approach for high contention programs to handle synchronization in shared-memory multiprocessors.

C.Ferri and other authors in [14] also compared locking with transaction based synchronization approach. They use frequency, power numbers and architectural assumptions based on simple cores for an embedded multiprocessor. According to their results transactional memory can provide clear performance advantages and also they prove the need of careful consideration to hardware design in order to meet the tight energy constraints of an embedded system. It has been shown by their graphs that transactional memory has advantages over locks in terms of ease of programming, performance and energy consumption. However, their applicability to embedded multi-core platforms has yet to be explored.

Energy efficient synchronization techniques for embedded systems like transactional memory and distributed semaphores were implemented by R. Iris Bahar and others are presented in [15]. According to their results embedded systems were constrained as compared to general purpose systems, and implementing these techniques using established mechanisms will not necessarily lead to an energy-efficient solution. To solve this issue, they developed an enhancement to the transactional hardware, which flushed the contents of the transactional cache back into the traditional cache hierarchy. There approach lead to a 17% savings in energy over a traditional transactional memory implementation. Although their mechanism was efficient in improving energy efficiency in contrast to locking, but no scheme till now is best in all benchmarks and in every situation.

We try to solve the issue of memory conflict in case of shared cache memory in MPSoC. Researchers have given many solutions like locking, transaction based synchronization [16] and many more but still there were drawback factors like high energy consumption, large cache miss rates, high CPU cycles etc.

Hence, to address these problems, we design a new synchronization technique which we are going to discuss in next section.

## 3. Proposed Solution

Synchronization conflicts cause transactions to abort and restart, causing the system to consume energy doing useless work [17]. Motivated by this tradeoff, in this paper we propose a semaphore solution with the intent of decreasing energy consumption and cache miss rate. In real time implementation, a semaphore is an integer variable accessed through two standard atomic operations: wait and signal. We can modify the value of semaphore but modifications to the integer value in the wait and signal operations must be executed atomically. Signal operation increases the semaphore count by one. Hence, when one processor modifies the semaphore value, no other processor can simultaneously modify that same semaphore value that is no other can access the same semaphore or data value on which semaphore is attached. While a processor is accessing a data item in shared memory, any other processor that tries to access the same must wait continuously to wait for the other processor to complete. However, in waiting, the processor can block itself. The block operation places a processor into a waiting queue associated with the semaphore attached with memory and the state of the processor is switched to the waiting state.

A blocked processor, waiting on a semaphore restarted when some other processor executes a signal operation. The processor is restarted by a wakeup operation, which changes the processor from the waiting state to the ready state. The processor is then placed in a ready queue. Our proposed algorithm as explained above is written below in fig 1. We have taken a counting semaphore initialized with value of maximum no. threads (processors) generated. A semaphore limits the number of concurrent users of a shared resource (memory) up to a maximum number. Threads can request access to the resource (decrementing the count of semaphore), and can signal that they have finished using the resource (incrementing the count of semaphore by signal operation) as we have done in our algorithm.

**Algorithm**

Step 1. Initialize the value of semaphore x with no. of processors for data values in shared memory.
Step 2. For each shared data value x
    For each processor accessing shared memory
        If x!=0 then processor can access the data
            and decrease the value of x by 1.
        Else wait and add requesting processor
            to waiting queue of semaphore.
Step 3. Whenever any processor completes execution, increments x by 1 i.e. signal for other processor and remove that processor from waiting queue and start execution.

Fig. 1 Proposed Algorithm

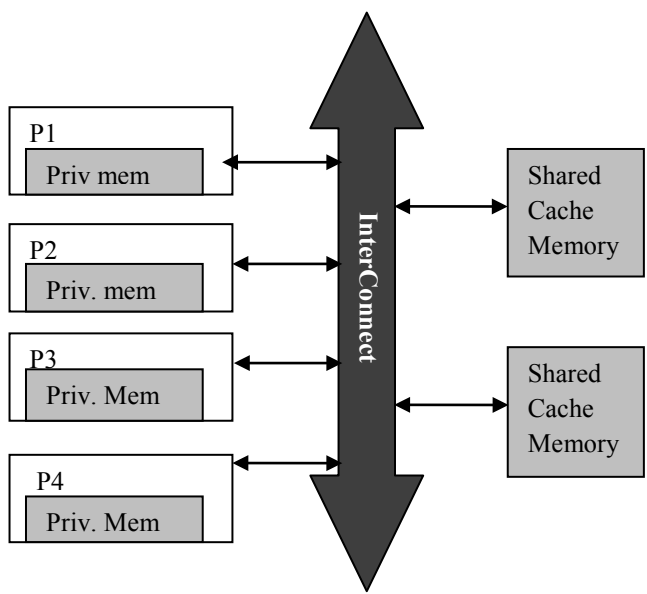

Fig.2 Implemented architecture

## 4. Simulation Environment

When the processor needs to read from or write to a location in main memory, it first checks whether a copy of that data is in the cache. If so, the processor immediately reads from or writes to the cache [18], [19], which is much faster than reading from or writing to main memory.

Caches used in systems can be divided into three types: [20] an instruction cache which can increase speed of executable instruction fetch, [21] a data cache to increase the speed of data fetch and store, and a translation look aside buffer (TLB) [22] to speed up virtual-to-physical address translation for both instructions and data. We have taken data cache and TLB cache both in consideration. The size of the cache line is usually larger than the size of the usual access requested by a CPU instruction [23], which ranges from 1 to 16 bytes. In a Multiprocessor System on chip (MPSoC) there can be two types of memories: Private memory for each processor and shared memories for processor [24].

We have taken up the scenario having four processors on a chip with private memories for each processor and two shared memories attached to each processor through a common ARM bus as shown in fig 2. We generate multiprocessing by using threads. The basic system configuration consists of a variable number of cores (each having direct-mapped D1 and T1 cache), a set of private memories (64 KB each), two shared memories (64KB). We use Linux operating system.

We developed our architecture on SimpleScalar functional simulator [25]. The SimpleScalar tool set provides a system software infrastructure to build modeling applications for analysis of program performance, detailed microarchitectural modeling and hardware-software co-verification. We can build modeling applications that simulate real programs running on a range of modern processors and systems in SimpleScalar. The tool set includes sample simulators ranging from a fast functional simulator (sim-fast) to a meticulous, dynamically scheduled processor model that supports non-blocking caches, speculative execution, and state-of-the-art branch prediction (sim-cache).

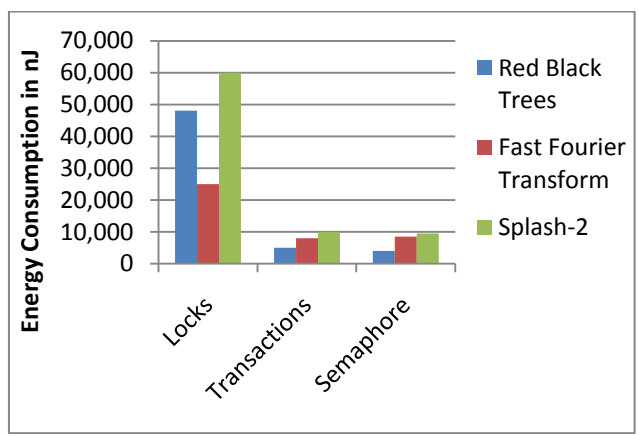

Fig. 3 Performance comparison in terms of energy consumption for locks,transactions vs semaphore

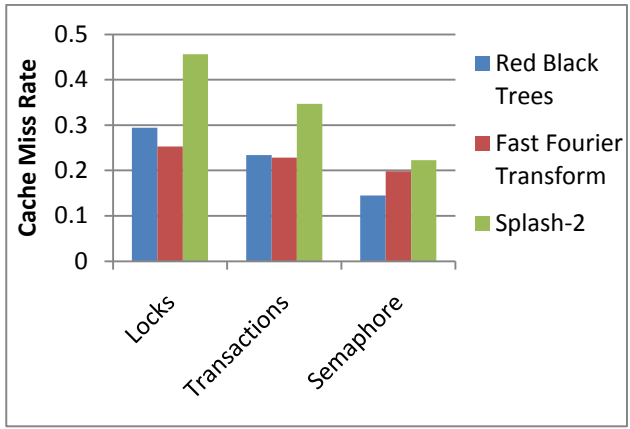

Fig. 4 Performance comparison in terms of cache miss rate for locks, transactions vs semaphore

We modify the original functional simulator sim-fast to calculate energy consumption [26]. We use sim-cache to evaluate cache miss rate for different benchmarks and techniques. We run simulation a number of times and final value is the average of all simulated values. We have chosen parallel application benchmarks as they require a large number of calculations and suffering from recurring conflict misses.

## 5. Experimental Results

This work is to compare locks and transactions to new proposed scheme semaphores in terms of energy consumption and cache miss rate. The bars labeled with locks in Fig 3 show the energy consumption of each of the benchmark runs in its initial configuration. The first three bars show memory accesses for the three benchmarks using locks while the next three bars are for the benchmarks using transactions, and the last three bars for our semaphores scheme. Shared memory accesses dominate energy consumption, making the locks simulation an unattractive energy-aware solution. Executing transactions mode consumes less energy than executing them with locks. Semaphores show almost comparable results of energy consumption to transactions.

The bars in Fig 4 show the performance of each of the benchmark runs reported above, in terms of cache miss rate. A shorter run displays better performance. For all benchmarks, transactions exceed locks, in terms of both energy and performance. Switching to semaphore execution mode improves both energy and performance. For this particular splash-2 microbenchmark, running transactions without semaphore and transaction execution mode generated more re-executions, which eventually resulted in more conflicts. These conflicts, in turn, led to more wasted instruction execution and cache miss rates and an increased number of cycles to complete the simulation.

## 6. Conclusion and Future Work

This paper proposes a new semaphore synchronization scheme and compares it with two other well-known shared memory synchronization techniques namely locks and Transaction. We compared these schemes based on energy consumption and cache miss rate using three benchmarks: Red Black Trees, Fast Fourier Transformation and a splash-2 microbenchmark as these are well known parallel applications.

We compared transactions and standard locks to our proposed semaphore, which require support from the OS. While locks can probably be optimized for energy in various ways (via sleep instructions, or platform-specific optimization), there is no standard way of modeling such optimizations, and no standard optimized locking package currently available to programmers. In this work, we do not claim to analyze the best of all conceivable lock implementations, but only to compare the standard, almost universally used locking libraries and transactions to semaphores. To summarize, for locks, the cost of synchronization depends on the number of locks in the application. For transactions, it d epends on the conflict rate. Therefore, transactional energy consumption depends on the specific scenario. For semaphore, performance depends on processor waiting time.

Our results prove that proposed semaphore synchronization technique have less cache miss rate in comparison to locks and transactions. In case of energy consumption, semaphore is almost comparable to transaction but much better than locks as shown by simulation results.

Our Future work consists of considering some other benchmarks and a wider range of architectural choices and hardware implementations.

**Shaily Mittal** received the B.Tech. degree in Computer Science and Engg. from Kurukshetra University, Kurukshetra, India in 2004 and M.Tech in computer science from M.D.U, Rohtak (India) in 2009. She was university ranker in her masters. She is currently pursuing her Phd from Jaypee University of Information & Technology, Waknaghat (India) from 2010. She is currently with ITM University, Gurgaon, India as Asst. Prof in CSE department. She had 5 years of teaching experience. She is author of 3 research papers during her masters. Her areas of interest include multiprocessor systems with a shared distributed memory subsystem and networks.

**Nitin** is Ex. Distinguished Adjunct Professor of Computer Science, University of Nebraska, Omaha, USA. Currently he is working as an Assistant Professor in the Department of Computer Science & Engineering and Information Technology, Jaypee University of Information Technology, Waknaghat, Solan, Himachal Pradesh, India.